\documentclass[11pt,twocolumn]{article}
\usepackage{float}
\usepackage[margin=1in]{geometry}

\usepackage{amsmath,amssymb,amsfonts}

\usepackage{newtxtext,newtxmath}   % cleaner Times-like font

\usepackage{booktabs}
\usepackage{array}
\usepackage{makecell}
\usepackage{multirow}

\usepackage{graphicx}
\usepackage{subcaption}

\usepackage[linesnumbered,ruled,vlined]{algorithm2e}
\usepackage{algorithmic}

\usepackage{authblk}

\usepackage[hyphens]{url}
\usepackage{hyperref}

\begin{document}



\title{Causally-Informed Reinforcement Learning for Adaptive Emotion-Aware Social Media Recommendation}

\author[1]{Bhavika Jain}
\author[1]{Robert Pitsko}
\author[1]{Ananya Drishti}
\author[1]{Mahfuza Farooque}

\affil[1]{Pennsylvania State University, State College, PA, USA \\

\texttt{bvj5274@psu.edu} \\
\texttt{rmp6022@psu.edu} \\
\texttt{azd6012@psu.edu} \\
\texttt{mff5187@psu.edu}}

\date{}   % leave empty for no date



%%  \includegraphics[width=\textwidth]{sampleteaser}
%%  \caption{Seattle Mariners at Spring Training, 2010.}
 %% \Description{Enjoying the baseball game from the third-base
%%  seats. Ichiro Suzuki preparing to bat.}
 %% \label{fig:teaser}

%%\received{20 February 2007}
%%\received[revised]{12 March 2009}
%%\received[accepted]{5 June 2009}

\maketitle
\begin{abstract}
Social media recommendation systems play a central role in shaping users' emotional experiences. However, most systems are optimized solely for engagement metrics, such as click rate, viewing time, or scrolling, without accounting for users' emotional states. Repeated exposure to emotionally charged content has been shown to negatively affect users' emotional well-being over time.

We propose an \textit{Emotion-aware Social Media Recommendation (ESMR)} framework that personalizes content based on users’ evolving emotional trajectories. ESMR integrates a Transformer-based emotion predictor with a hybrid recommendation policy: a LightGBM model for engagement during stable periods and a reinforcement learning agent with causally informed rewards when negative emotional states persist. Through behaviorally grounded evaluation over 30-day interaction traces, ESMR demonstrates improved emotional recovery, reduced volatility, and strong engagement retention. ESMR offers a path toward emotionally aware recommendations without compromising engagement performance.
\end{abstract}


\section{Introduction}

Social media platforms wield immense influence over users’ emotional experiences, yet their underlying recommender systems are typically optimized for behavioral engagement alone, prioritizing clicks, views, and watch time without regard for emotional impact. This mismatch has raised growing concerns about the role of algorithmic content curation in reinforcing stress, anxiety, or affective volatility~\cite{algobad_2, algobad, ytDoc}.

Prior emotion-aware recommenders, such as EARS~\cite{qianEARS}, typically use static mood-alignment strategies that match content to a user's short-term emotions. However, these systems lack mechanisms to address persistent negative emotions, which involve repeated exposure to high-intensity negative states (e.g., stress, disappointment) over multiple sessions. Likewise, reinforcement learning approaches such as RL4Rec and list-wise RL frameworks~\cite{zhao2018deep, Lin2025} focus on engagement maximization without explicit emotion-state triggers or emotion transition modeling. These systems typically treat all user states uniformly rather than adapting policies based on emotional volatility—that is, rapid or oscillating shifts between conflicting emotional states~\cite{norealdata}. In contrast, ESMR dynamically activates RL only when users enter a prolonged vulnerable state and shape recovery using causally linked features, enabling personalized and emotionally adaptive trajectories.

In this work, we propose Emotion-State-aware Social Media Recommendation (ESMR), a hybrid framework that integrates LightGBM -based engagement scoring~\cite{Ke2017LightGBM}, reinforcement learning for affective recovery~\cite{zhao2018deep, Lin2025}, and causal modeling for emotion transition shaping. ESMR selectively activates reinforcement learning (RL) when users enter prolonged negative emotional states, guiding content delivery that balances engagement with emotional recovery. 

Emotional states are inferred through clustering over engagement signals such as scrolling time, video viewing behavior, and interaction frequency, inspired by prior affect-behavior correlations~\cite{gao2017behavioral, zhang2021emotionaware}. Our goal is to move beyond affect-agnostic recommenders by prioritizing user well-being and designing emotionally intelligent systems that promote recovery, stability, and long-term resilience~\cite{bonner2019causalembeddingsrecommendationextended}. 
\section{Background and Related Works}
This section reviews foundational work in emotion-aware recommendation systems, reinforcement learning in recommendation, and causal discovery in affective modeling. We discuss existing methods for integrating user emotional feedback, highlight the limitations of current engagement-centric systems, and present how our approach addresses these gaps through hybrid RL and emotion shaping.

\subsection{Emotion-Aware Recommender Systems}
Emotion-aware recommenders aim to incorporate users’ psychological and affective states into the personalization pipeline. Models like EARS~\cite{qianEARS} map user behaviors to mood states and use emotionally labeled content to guide recommendations. While effective in aligning short-term mood, such approaches are generally reactive and do not support long-term adaptation or emotional recovery.

Most systems also assume a direct correlation between content valence and user mood without distinguishing the causal drivers of emotional shifts. This can result in superficial personalization, where emotionally salient content is overexposed, potentially reinforcing affective volatility and worsening users' emotional trajectories over time. Empirical studies have shown that frequent exposure to emotionally charged content (e.g., on platforms like Facebook) is associated with significant declines in subjective well-being and mood stability~\cite{algobad_2}. A recent research shows possible interventions to improve emotional states and user well-being~\cite{rec_goodwork}.

\subsection{Reinforcement Learning in Recommendation}

Reinforcement Learning (RL) has become a powerful tool for adaptive personalization in sequential recommendation tasks. List-wise RL approaches, such as Deep Reinforcement Learning for List-wise Recommendations (DRR)~\cite{zhao2019deepreinforcementlearninglistwise}, model the recommendation process as a Markov Decision Process (MDP), where \( s_t \) represents user state (including emotional and behavioral features) \( a_t \) is the recommended content, and \( r_t \) is the reward, typically capturing long-term user utility. These models optimize ranked lists to maximize cumulative rewards over time. Policy-gradient methods have also been applied to recommender systems, directly optimizing ranking metrics via pairwise policy gradient~\cite{rl_policy_intro}.

Conversational and contextual bandit models have been explored to enhance online interactivity and exploration, allowing systems to adapt more quickly to evolving user preferences through real-time feedback~\cite{conversational}.

Despite their success in engagement optimization, most RL-based recommenders focus on engagement-centric rewards, such as clicks, dwell time, or views, often neglecting user-centric outcomes like satisfaction or emotional stability. This narrow objective can lead to feedback loops, where sensational or emotionally provocative content dominates, undermining long-term user well-being~\cite{Fairness}.

\subsection{Causal Modeling for Emotion Prediction and Reward Shaping}

To move beyond correlation-based personalization, causal inference has emerged as a principled framework for identifying the true effects of content or actions on user outcomes. In recommender systems, this shift has been exemplified by methods such as Causal Embeddings, which aim to disentangle causation from mere association using treatment-aware representations~\cite{Bonner2017CausalEmbed}. More broadly, causal modeling has been applied to infer how recommendations influence engagement trajectories and long-term utility~\cite{gao2023causal}.

In emotion modeling, causal reasoning helps distinguish between content that actively induces emotional transitions and content that merely co-occurs with them. We extend this by learning user-specific causal graphs over behavioral and content features, identifying top causal parents \( \mathcal{P} \) of next-day emotional states. Using structure learning algorithms like DirectLiNGAM~\cite{directlingam}, we recover directed acyclic graphs (DAGs) that capture individual-level emotional dependencies (Section~\ref{sec:causal_dag}). While DirectLiNGAM offers scalability, its assumptions (e.g., linearity, no unobserved confounders) may not fully hold in real-world user behavior; future work will explore alternatives such as PC and NOTEARS. Under Pearl’s framework, these DAGs enable inference of latent emotional states via backdoor adjustment or surrogate conditioning~\cite{pearl2009causality}, aligned with causal representation learning efforts~\cite{louizos2017causal}. We use the top causal parents—features with the strongest influence on emotional shifts—to shape the RL reward function, promoting actions that are both engaging and emotionally beneficial.

\section{Methodology}
\begin{figure}[htbp]   % h = here, t = top, b = bottom, p = page of floats
    \centering
    \includegraphics[width=\linewidth]{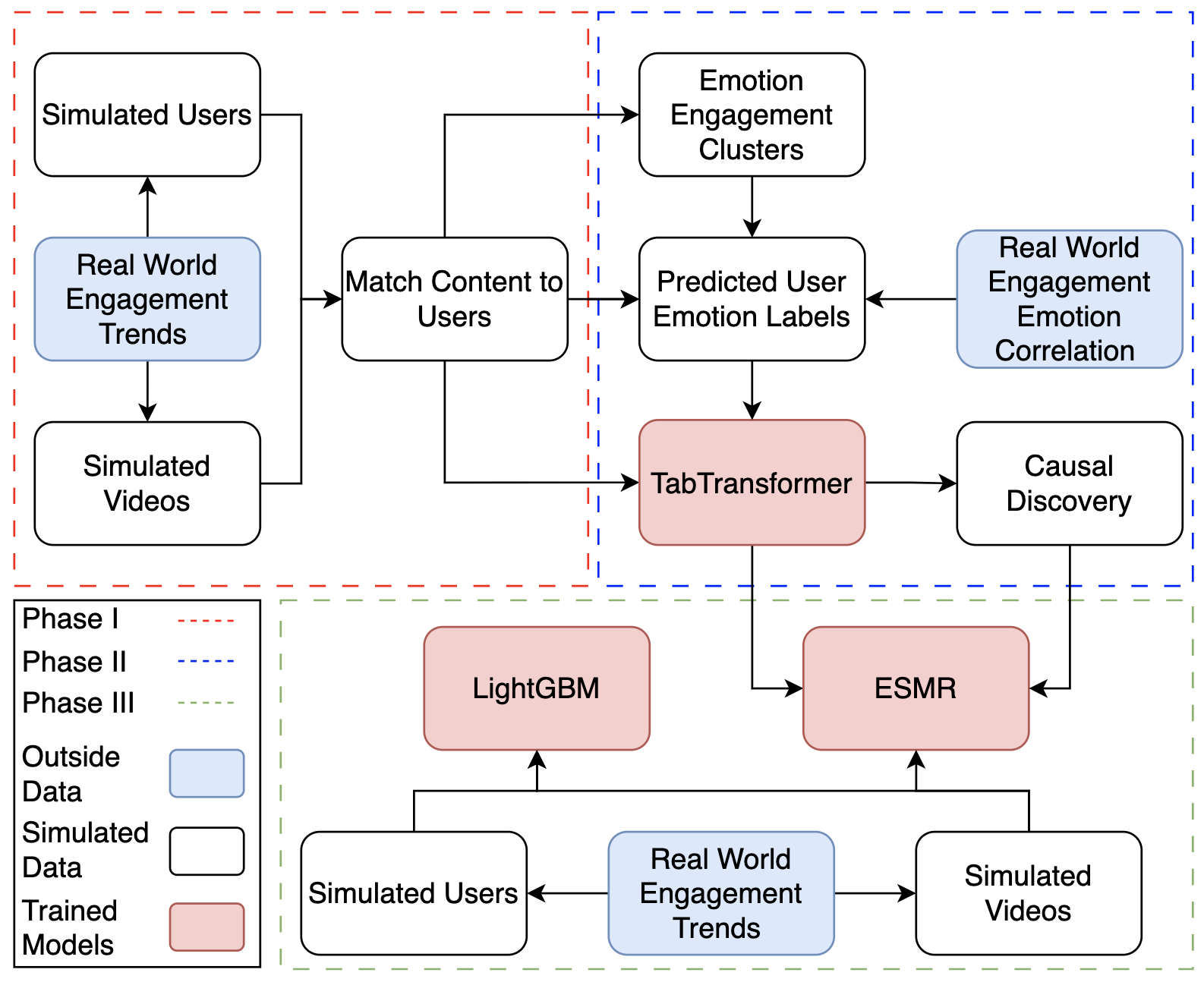}
    \caption{Overview of our methods. Phases I–II simulate users and predict emotional states; Phase III implements ESMR, a hybrid LightGBM–RL model with causal reward shaping based on emotion-influencing features.}
    \label{fig:method2}
\end{figure}

Our work consists of three phases (Figure~\ref{fig:method2}). Phase I simulates content and users (Section~\ref{sec:phase-I}), Phase II trains an emotion prediction model and discovers causal relationships (Section~\ref{sec:phase-II}), and Phase III implements ESMR (Section~\ref{sec:phase-III}).

\subsection{Phase I: Data Simulation across time}
\label{sec:phase-I}

\subsubsection{\textbf{Content Simulation and Emotion Labeling}}
\label{sec:content_simulation}
% how each video is created with emotion tags, categories, themes, etc.
To create a realistic social media environment for model testing, we simulate 1,000 videos, each labeled with emotion, duration, theme, category, and engagement potential, ensuring variability across content types and emotional appeal.

\textbf{Theme and Category.} Videos are assigned to one of four categories: \textit{educational}~\cite{enter_edu_insp}, \textit{entertainment}~\cite{enter_edu_insp}, \textit{news}~\cite{news_enter_informative}, or \textit{inspirational}~\cite{enter_edu_insp}, with themes reflecting trending topics such as \textit{mental health}~\cite{political_mental}, \textit{politics}~\cite{political_mental}, \textit{motivation}~\cite{lonliness_stress_motivation}, and \textit{relationships}~\cite{relationships}, which align with current real-world social media content.

\textbf{Duration.}Video durations are drawn from a clipped Gaussian distribution ($\mu = 30$, $\sigma = 15$), bounded between 10–90 seconds, reflecting the typical duration of social media videos on platforms like TikTok and Instagram~\cite{formulation-skipped}.

\textbf{Emotion Assignment and Intensity.} Each video is labeled with one of eight affective states (e.g., \textit{happy}, \textit{anxious}, \textit{stressed}), drawn from an imbalanced categorical distribution to reflect the emotional biases observed in social media content, with emotions like \textit{excited}, \textit{happy}, and \textit{anxious} having higher sampling probabilities~\cite{virality_vids}. Intensity scores (1–10) are drawn from emotion-specific Gaussian distributions~\cite{intenisty_vid_emotion}, grounded in the valence-arousal model of emotion~\cite{valence_arousal}.

\textbf{Video Engagement Score.} Each video receives a composite engagement score, computed from duration, category-level base engagement~\cite{cat_specific_engag}, emotional intensity, and virality multipliers~\cite{popular_engag, virality_vids}. These engagement scores approximate the likelihood of user interaction on platforms and serve as both predictive features (Section \ref{sec:phase-II}) and reward signals in reinforcement learning (Section \ref{sec:phase-III}). The distribution of video engagement scores varies across emotional labels. Highly arousing emotions such as \textit{excited} lead to greater variability in engagement scores, while more neutral emotions like \textit{disappointed} tend to show lower engagement scores and narrower distributions. This reflects the underlying assumption that different emotional states influence varying levels of user interaction with content, where more emotionally engaging content elicits higher interaction variability and neutral or negative content elicits lower and more stable interaction \cite{zhao2021emotionengagement, wang2012emotional,gao2017behavioral}.

\subsubsection{\textbf{User Modeling and Engagement Simulation}}
\label{sec:user_eng}
% how user profiles are constructed
% what engagement signals (scrolling, watching, liking) are generated
% how emotion transitions work per content/day
We simulate 1,000 users over 30 days; each assigned a behavioral profile $p_u \in \{\textit{casual}, \textit{engaged}, \textit{highly active}\}$~\cite{relationships, user_cat}, which governs daily session patterns and engagement intensity. On day $t$, user $u$ logs scrolling time ($S_u^{(t)}$), watch duration ($V_u^{(t)}$), login frequency ($L_u^{(t)}$), interaction features, and churn status ($z_u^{(t)}$). These metrics are drawn from empirically grounded distributions reflecting real-world social media behaviors.

\textbf{Session Behavior.}
($S_u^{(t)}$) and ($V_u^{(t)}$) are calculated in Equation \ref{eq:scorlling_time_vid_watching}.
\begin{equation}
\label{eq:scorlling_time_vid_watching}
S_u^{(t)} 
\sim \Gamma(k_p, \theta_p), \quad V_u^{(t)} \sim \mathcal{U}(a_p, b_p) \cdot M^{(t)}
\end{equation}

\(k_p, \theta_p, a_p, b_p\) are shape and scale parameters~\cite{gamma_scrolling_2} , determined by user profile \(p_u\), and \(M^{(t)}\) models temporal spikes in Equation \ref{eq:temporal_spikes}.

\begingroup
\begin{equation}
\label{eq:temporal_spikes}
M^{(t)} = 
\begin{cases}
\text{Uniform}(1.2, 1.8) & \text{if} \, t \in \mathcal{T}_{\text{spike}} \\
1.2 & \text{if weekend} \\
1.0 & \text{otherwise}
\end{cases}
\end{equation}
\endgroup

\(\mathcal{T}_{\text{spike}}\) denotes randomly sampled high-engagement days that mimic virality or usage surges~\cite{likes-more, weekend_spikes}. 

\textbf{Video Skipping.}
Each user skips a fraction of the encountered videos, which is modeled probabilistically~\cite{random-skipped}. Let \( \bar{v} = 45 \)s denote the average video duration, and define the approximate number of videos encountered as \( n = V_u^{(t)} / \bar{v} \). \( n_s \) skipped videos, \( n_w \) fully watched videos~\cite{random-skipped, formulation-skipped} are partitioned in Equation \ref{eq:skip_vids}.

\begin{equation}
\label{eq:skip_vids}
n_s = \lfloor 0.4n \rfloor, \quad n_w = n - n_s
\end{equation}

The effective video-watching time is adjusted in Equation \ref{eq:adjust_vid_watch_time_skip}.

\begin{equation}
\label{eq:adjust_vid_watch_time_skip}
V_u^{(t)} = \bar{v} \cdot n_w + \sum_{i=1}^{n_s} \gamma_i \cdot \bar{v}, \quad \gamma_i \sim \mathcal{U}(0.2, 0.7)
\end{equation}

\( \gamma_i \) captures the partial duration of the \( i^{\text{th}} \) skipped video.

\subsubsection*{\textbf{Churn and Re-engagement}}

User churn is modeled using a smoothed engagement signal in Equation \ref{eq:churn}

\begin{equation}
\label{eq:churn}
\Delta E_u^{(t)} = 0.8 \cdot \Delta E_u^{(t-1)} + 0.2 \cdot \operatorname{sign}(E_u^{(t)} - \tau)
\end{equation}

\( E_u^{(t)} \) is the composite engagement score on day \( t \), and \( \tau \) is a global threshold controlling disengagement sensitivity. The smoothed signal \( \Delta E_u^{(t)} \) captures short-term engagement trends while filtering volatility. 

We define the binary churn indicator in Equation \ref{eq:z_status}.

\begin{equation}
\label{eq:z_status}
z_u^{(t)} \sim \operatorname{Bernoulli}(\phi_u^{(t)}), \quad \phi_u^{(t)} = \sigma\left(p_u + \alpha \cdot \Delta E_u^{(t)}\right)
\end{equation}

\( z_u^{(t)} = 1 \) indicates that user \( u \) has churned on day \( t \), and \( \sigma(\cdot) \) denotes the sigmoid function. The churn probability \( \phi_u^{(t)} \in (0, 1) \) is determined by a personalized base propensity \( p_u \in [0, 1] \), sampled from the user profile and adjusted by recent changes in engagement via \( \Delta E_u^{(t)} \). To reflect behavioral volatility, churn is modeled stochastically: users who churn may re-engage on future days with a small fixed probability (e.g., 0.1), enabling realistic cycles of disengagement and return~\cite{churn}.

\textbf{Content Preferences.} User preferences over content categories are modeled using a Dirichlet prior~\cite{gamma_scrolling_2}. Specifically, we assume
\[
c_u^{(t)} \sim \mathrm{Dirichlet}(\alpha),
\]
where $c_u^{(t)} \in \mathbf{R}^K$ denotes the preference vector over $K=4$ content categories. The concentration parameter $\alpha$ is user profile–dependent: casual users tend to exhibit more uniform preferences, while engaged users typically allocate higher probability mass to certain favored categories. The resulting preference vector is then used to drive the content engagement dynamics formalized in Equation~\ref{eq:content_pref}.

\begin{equation}
\label{eq:content_pref}
\quad \vec{E}_u^{(t)} = V_u^{(t)} \cdot \vec{c}_u^{(t)}
\end{equation}

\( \vec{E}_u^{(t)} \) allocates \( V_u^{(t)} \) across \(K = 4\). This distribution is normalized so that \( \sum_{i=1}^{K} E_{u,i}^{(t)} = V_u^{(t)} \),  ensuring the total engagement aligns with the user's daily viewing time.

\textbf{Interaction Behavior.}
User interactions—including \(L_u^{(t)}\), posts, likes, comments, and shares—are simulated using profile-dependent distributions: Poisson for logins\cite{poisson-login}, bursty post-event activity for posts\cite{neg-binom}, a long-tailed distribution approximated by a lognormal for likes\cite{lognormal}, and empirical distributions for comments and shares\cite{sharing}. These signals provide fine-grained behavioral context for downstream modeling.

\subsubsection{\textbf{Daily Log Generation and Feature Construction}}
\label{sec:assign_vids}
% how all this is structured into the dataset you use: user-day records
% what features go into the model (video metadata, engagement, emotion scores, etc.)

Each execution of Algorithm~\ref{alg:daily-log} produces a structured daily record \( \mathcal{D}_u^{(t)} \) containing assigned videos \( \mathcal{V}_{\text{assign}}^{(t)} \), skipped content \( \mathcal{V}_{\text{skip}}^{(t)} \), and derived features such as composite engagement \( E_u^{(t)} \), smoothed change \( \Delta E_u^{(t)} \), scroll-to-watch ratio \( \rho_u^{(t)} \), and category-level time allocation \( \vec{\tau}_u^{(t)} \). Content selection follows a 70–30 rule: 70\% from top engagement scores, 30\% randomly sampled to simulate novelty exposure~\cite{virality_vids,70-30_split}.

The final dataset includes 30,000 user–day records spanning 1,000 users over 30 days, each with over 25 engineered features. To validate behavioral realism, we examine feature correlations: engagement \( E_u^{(t)} \) correlates strongly with likes (\( \rho = 0.66 \)), moderately with comments (\( \rho = 0.43 \)), and weakly with shares (\( \rho = 0.29 \)), reflecting effortful interaction tiers seen on social platforms~\cite{sharing}. This hierarchical trend supports the plausibility of simulated user behavior.

\begin{algorithm}[htbp]
\caption{Daily Record Generation for User $u$ on Day $t$}
\label{alg:daily-log}
\KwIn{User profile $p_u$, content pool $\mathcal{C}$, recent interaction history}
\KwOut{Structured record $\mathcal{D}_u^{(t)}$ with assigned content and engagement features}

\If{$z_u^{(t)} = 1$}{\textbf{return} $\emptyset$ \tcp*[f]{User has churned}}

Sample available watch time \( V_u^{(t)} \sim \text{Gamma}(p_u) \)\;

Draw content preference vector \( \vec{c}_u^{(t)} \sim \text{Dirichlet}(\vec{\alpha}) \)\;

Initialize sets: \( \mathcal{V}_{\text{assign}}^{(t)} \leftarrow \emptyset \), \( \mathcal{V}_{\text{skip}}^{(t)} \leftarrow \emptyset \), time used \( \leftarrow 0 \)\;

\ForEach{category \( k \in \mathcal{K} \)}{
    Retrieve candidates \( \mathcal{C}_k \subset \mathcal{C} \) based on \( \vec{c}_u^{(t)} \)\;
    Apply 70-30 sampling: 70\% from top-engagement, 30\% random\;
    Filter out videos seen in the past 5 days\;

    \ForEach{video \( v \in \mathcal{C}_k \)}{
        Compute skip probability: \( P_{\text{skip}}(v) \propto 1 - \text{engagement}(v) \)\;
        Sample outcome (skip vs. watch)\;
        \uIf{skipped}{Add \( v \) to \( \mathcal{V}_{\text{skip}}^{(t)} \)}
        \Else{Add \( v \) with timestamp to \( \mathcal{V}_{\text{assign}}^{(t)} \)}
        Update time used; \If{limit exceeded}{\textbf{break}}
    }
}

Compute user features: \\
\quad -- \( E_u^{(t)} \): composite engagement score \\
\quad -- \( \Delta E_u^{(t)} \): smoothed daily engagement change \\
\quad -- \( \rho_u^{(t)} \): scroll-to-watch ratio \\
\quad -- \( \vec{\tau}_u^{(t)} \): category time distribution

\textbf{return} \( \mathcal{D}_u^{(t)} = \{ \mathcal{V}_{\text{assign}}^{(t)}, \mathcal{V}_{\text{skip}}^{(t)}, E_u^{(t)}, \Delta E_u^{(t)}, \rho_u^{(t)}, \vec{\tau}_u^{(t)}, \text{timestamps} \} \)
\end{algorithm}

\subsection{Phase II: User Emotion Prediction and Causal Discovery}
\label{sec:phase-II}

In Phase II, we infer daily user emotions from engagement signals using cluster-informed labels and a TabTransformer classifier, bridging the gap between latent affect and observed behavior.

\subsubsection{\textbf{Clustering}}
\label{sec:clustering}
We apply K-Means clustering ($k=5$), selected via the elbow method, to standardized daily records \( \mathcal{D}_u^{(t)} \) constructed in Section~\ref{sec:assign_vids}. We further refine labels using rule-based conditions: users exhibiting complete inactivity are marked as \textit{churned}, while sharp drops in engagement trigger a \textit{stressed} label~\cite{doomscroll, churn}.
Each cluster is heuristically mapped to an emotion based on dominant traits, grounded in behavioral literature~\cite{random-skipped, valence_arousal}. These cluster-derived labels are summarized in Table~\ref{tab:cluster-mapping}. 

\begin{table}[htbp]
\setlength{\textfloatsep}{6pt plus 1pt minus 1pt}
\centering
\caption{Cluster-to-emotion mapping based on engagement traits.}
\label{tab:cluster-mapping}
\begin{tabular}{c p{0.46\linewidth} l}
\toprule
\textbf{Cluster} & \textbf{Behavioral Pattern} & \textbf{Emotion Label} \\
\midrule
0 & High interaction, consistent viewing & Happy~\cite{happy} \\
1 & No activity across features & Churned~\cite{another_churn, churn} \\
2 & Passive use, low engagement & Disappointed~\cite{emotion_disappointed} \\
3 & Burst in likes and shares & Excited~\cite{excited_emotion} \\
4 & High scroll, low watch time & Stressed~\cite{doomscroll} \\
\bottomrule
\end{tabular}
\end{table}

Although cluster labels are noisy approximations of true emotions, their alignment with known behavioral signatures enables their use as training targets for supervised classification \cite{gao2017behavioral}. 

% We omit boxplot and variance plots due to space; Add github

\subsubsection{\textbf{Transformer-based Emotion Prediction}}
\label{sec:transformers}

\begin{figure}[h]
    \centering
    \includegraphics[width=\linewidth]{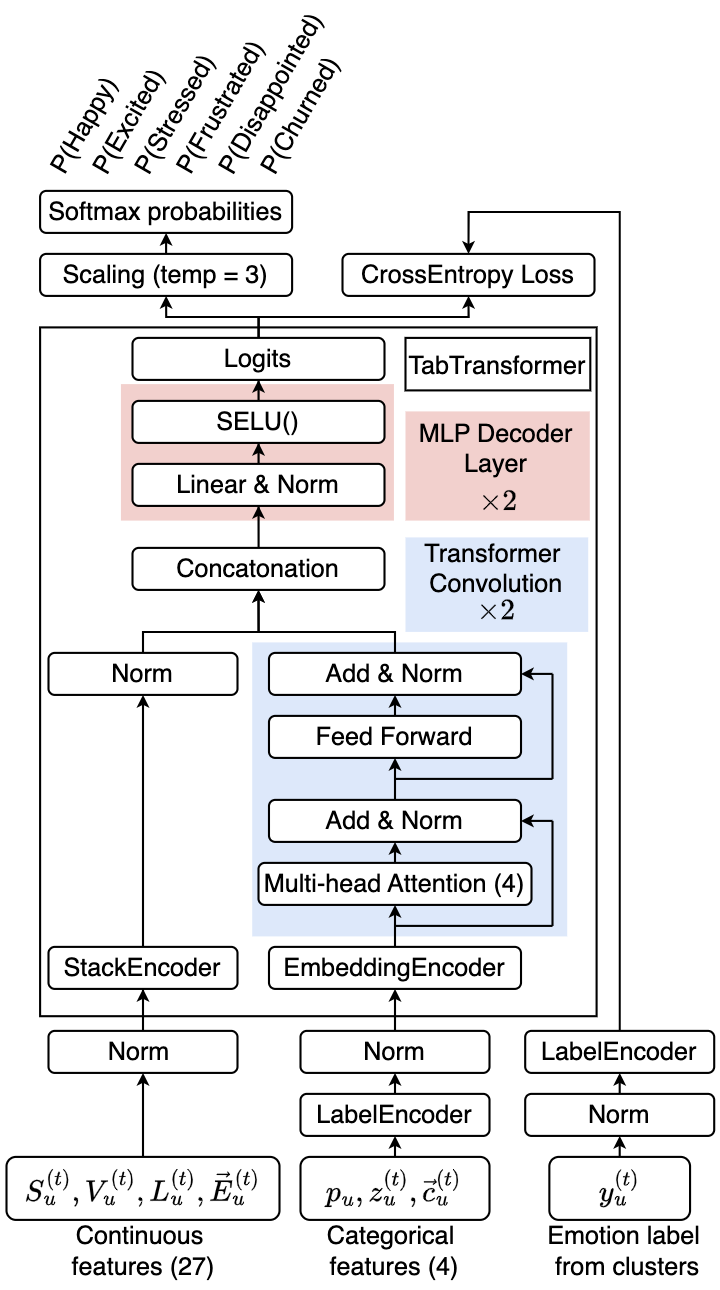}
    \caption{Architecture of TabTransformer.}
    \label{fig:TabSize}
\end{figure}

We adopt TabTransformer~\cite{tabtransformer} for user emotion classification from tabular engagement data. Its hybrid design, combining column-wise attention over categorical embeddings with MLP layers, is well-suited to our heterogeneous feature space (see Figure~\ref{fig:TabSize}). Inputs include 27 continuous features and four categorical fields, capturing user profiles, content attributes, and prior emotional context.

The model outputs softmax probabilities over six emotion classes and is trained with cross-entropy loss. TabTransformer showed stronger generalization across user days and handled sparse behavioral signals more robustly than FT-Transformer~\cite{ft}, TabNet~\cite{tabnet}, and Time-Series Transformers~\cite{timeseries}. The alternatives struggled with overfitting (FT), unstable training (TabNet), or limited temporal signal (Time-Series). The results for emotion prediction can be seen in Section~\ref{sec:emotion_prediction_results}

\subsubsection{\textbf{Causal Discovery for Emotion Transitions}}
\label{sec:causal}
While supervised models can predict emotional states from engagement signals, they fail to explain why transitions occur. To uncover behavioral drivers of affective change, we apply causal discovery on user–day engagement logs. Each user’s daily record is flattened into a temporal dataset containing both behavioral features and next-day emotional outcomes (e.g., valence change, next-day stress).

We adopt \textit{DirectLiNGAM}~\cite{directlingam}, a linear non-Gaussian algorithm for learning directed acyclic graphs (DAGs), which is more robust than alternatives like NOTEARS~\cite{dag} in sparse, non-Gaussian behavioral data. DAGs are learned over 19 features, including engagement scores, scrolling/watching time, and emotion intensities.

For each emotional outcome, we extract the full set of its causal parents. These features form the set \( \mathcal{P} \) used in the causal reward component \( R_{\text{cause}}(s, a) \), enabling the RL agent to assign credit to improvements in causally grounded variables. If a user is in a vulnerable state (e.g., stressed or disappointed), we amplify causal bonuses to prioritize recovery (Refer Section \ref{sec:rl}). This structure improves both interpretability and alignment between the reward signal and emotional stabilization goals (Refer Section \ref{sec:ablation}).

%Rewards are boosted when actions influence features that causally impact positive emotional outcomes or reduce drivers of negative transitions. This enables our reinforcement learning agent to make emotionally beneficial decisions even in the absence of explicit engagement gains—a hallmark of our ESMR strategy.
%
\begin{figure}[htbp]
  \centering
    \includegraphics[width=0.98\columnwidth]{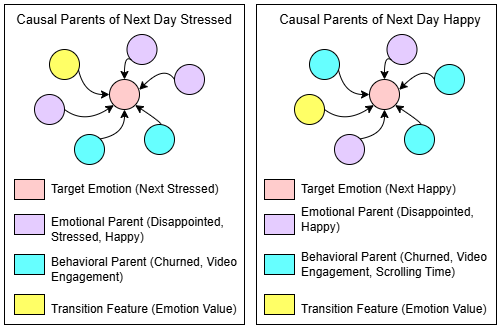}
    \caption{DAG of Causal Parents for the Happy and Stressed State, highlighting emotional, behavioral, and transitional features identified through causal discovery. Color-coded nodes reveal key drivers of happiness and stress based on user engagement dynamics.}
    \label{fig:causal_parents}
\end{figure}
\begin{table}[htbp]
\setlength{\textfloatsep}{6pt plus 1pt minus 1pt}  % Safe spacing
\centering
\small
\caption{Top-3 causal parents per emotional outcome from the learned DAG. Signs indicate directionality. Full parent sets are used in reward shaping.}
\label{tab:causal_parents_summary}
\begin{tabular}{p{0.34\linewidth} p{0.60\linewidth}}  % Adjusted widths
\hline
\textbf{Emotional Target} & \textbf{Top Causal Parents (Name + Sign)} \\
\hline
\textit{Next-Day Stressed} & \( V_u^{(t)} \) (+), \( \Delta E_u^{(t)} \) (–), Intensity (–) \\
\textit{Valence Change} & \( \Delta E_u^{(t)} \) (+), Score (–), \( \Delta \text{Intensity} \) (+) \\
\textit{Next-Day Happy} & \( V_u^{(t)} \) (+), \( S_u^{(t)} \) (+), \( \Delta E_u^{(t)} \) (+) \\
\hline
\end{tabular}
\end{table}

To support interpretability, Table~\ref{tab:causal_parents_summary} presents the top 3 strongest causal parents for three core emotional targets. The full parent sets are used in training.

\subsection{Phase III: Emotion-aware Social Media Recommender (ESMR) via Hybrid Policy}
\label{sec:phase-III}

To balance engagement with long-term emotional stability, we implement a hybrid recommendation policy
\(\pi_t \in \{\text{LightGBM}, \text{RL}\}\) alternating between supervised engagement-based recommendations (LightGBM) and reinforcement learning (RL)-based personalization. 

At the start of each day \( t \), the system computes the user’s current state \( s_t \). This state is derived from recent behavioral signals, emotional states (e.g., \( y_u^{(t)} \)), and content interaction history \( \mathcal{D}_u^{(t-1)} \).

The 3-day threshold is empirically supported by the smoothed emotion trajectories shown in Figure~\ref{fig:emotion-trend}, where emotional shifts often stabilize after a short-term persistence period. This hybrid switching mechanism draws inspiration from adaptive RL policy strategies used in sequential recommendation tasks~\cite{rl_intro, rl-intro}.

\textbf{Notation.} Let \( s_t \) denote the user’s state at day \( t \), incorporating recent emotional intensities, engagement features, and content interaction trends. The active policy \( \pi_t \) selects an action \( a_t \in \mathcal{A} \), corresponding to a video recommendation \( v_{a_t} \). After observing the user’s behavioral and emotional response, the system computes a scalar reward \( r_t \), predicts the next emotional state \( e_t \) using the TabTransformer, and logs the tuple \( (s_t, a_t, r_t, e_t, \pi_t) \) for learning and evaluation. The RL agent maintains a Q-function \( Q(s_t, a_t) \), as in standard MDP-based RL~\cite{rlbook}, and receives a composite reward function \( R(s_t, a_t) \), which consists of the engagement-based reward \( R_{\text{eng}}(s_t, a_t) \), the emotion-alignment reward \( R_{\text{emo}}(s_t, a_t) \), and a causal shaping bonus \( R_{\text{cause}}(s_t, a_t) \) where the causal bonus uses a subset of user features \( \mathcal{P} \subset \mathcal{F} \), selected from the top-ranked causal parents of emotional targets discovered in Section~\ref{sec:causal}~\cite{madumal2020explainable}. These causal features help align rewards with emotionally stabilizing behavior by encouraging improvements in behaviorally meaningful signals.

\begin{figure*}[htbp]
  \centering
  \includegraphics[width=\textwidth]{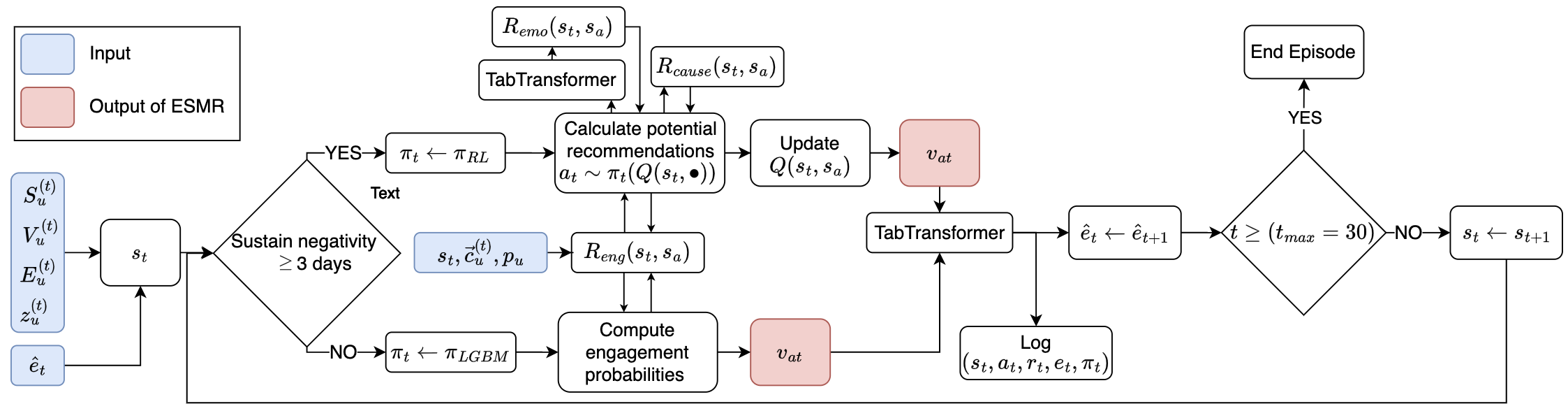}
  \caption{Symbolic execution flow of the ESMR hybrid framework. At each day \(t\), state \(s_t\) and emotion history determine the active policy \(\pi_t\). Actions \(a_t\) are selected using LightGBM (during stable periods) or RL (during vulnerable states). Rewards \(r_t\) and emotional states \(\hat{e}_t\) are updated accordingly and logged for learning and evaluation.}
  \label{fig:phase3}
\end{figure*}

The full recommendation process operates as a Markov Decision Process (MDP) defined by state space \( \mathcal{S} \), action space \( \mathcal{A} \), reward function \( R \), and discount factor \( \gamma \), with transitions governed by both engagement behavior and affective feedback~\cite{rlbook}. For a symbolic overview of the ESMR workflow, see Figure~\ref{fig:phase3}.

\subsubsection{\textbf{LightGBM-Based Engagement Modeling}}
\label{sec:gbm}

When \( \pi_t = \text{LightGBM} \), the system generates recommendations using LightGBM \cite{Ke2017LightGBM}, trained to predict user engagement from content features and behavioral history. Each training sample corresponds to a user–video–day tuple, with features such as \( S_u^{(t)} \), interaction trends, content metadata, and \( \hat{y}_u^{(t)} \) from Section~\ref{sec:phase-II}. Labels are binary, indicating whether the video was watched or skipped. The model is trained on 30 days of simulated data using GPU-accelerated LightGBM, with early stopping based on AUC (stops on 0.81).

During inference, LightGBM ranks 1,000 candidate videos per user–day by predicted engagement probability. The top-1 video \(v_{a_t}\) is served as the final recommendation. While the model does not adapt online, its predictions guide high-throughput engagement optimization during emotionally stable periods, contributing to the engagement reward term \(R_{\text{eng}}(s_t, a_t)\) for downstream evaluation.

\subsubsection{\textbf{Reinforcement Learning Agent for Emotion-Aware Recommendations}}
\label{sec:rl}

The reinforcement learning (RL) policy of ESMR is activated when the user enters a vulnerable state, defined as three or more consecutive days dominated by negative or emotionally neutral affect. In such cases, the policy \(\pi_t\) switches to RL, temporarily overriding LightGBM to promote emotional recovery. The agent is trained using episodic tabular Q-learning~\cite{sutton2018reinforcement} to recommend emotionally aligned content and maximize personalized rewards.

\textbf{Reward Function Design.}  
The total reward \( R(s_t, a_t) \) integrates three components: engagement \( R_{\text{eng}}(s_t, a_t) \), emotional recovery \( R_{\text{emo}}(s_t, a_t) \), and causal feature shifts \( R_{\text{cause}}(s_t, a_t) \). The engagement term \( R_{\text{eng}}(s_t, a_t) \) captures immediate user response, such as watch time \( V_u^{(t)} \) and scroll depth \( S_u^{(t)} \), normalized using the hyperbolic tangent function to avoid reward inflation. The emotion-based term \( R_{\text{emo}}(s_t, a_t) \) assigns a fixed bonus \( \eta \) when the user transitions from a negative to a neutral or positive emotional state \( y_u^{(t)} \). Finally, the causal shaping term \( R_{\text{cause}}(s_t, a_t) \) aggregates weighted improvements in key causal parent features \( \mathcal{P} \), identified through DirectLiNGAM (see Section~\ref{sec:causal})~\cite{madumal2020explainable}. The overall reward is computed as shown in Equation~\ref{eq:shaped-reward}.

\begin{equation}
\label{eq:shaped-reward}
R(s_t, a_t) = R_{\text{eng}}(s_t, a_t) + \alpha \cdot R_{\text{emo}}(s_t, a_t) + \beta \cdot R_{\text{cause}}(s_t, a_t)
\end{equation}

This reward is operationalized in Algorithm~\ref{alg:reward-computation}, where \(\lambda_f\) denotes the weight for each causal parent feature \(f \in \mathcal{P}\).

\begin{algorithm}[htbp]
\caption{Reward Computation in Emotion-Aware RL}
\label{alg:reward-computation}
\KwIn{State $s$, Action $a$, Causal Parents $\mathcal{C}$, Emotion state $e$}
\KwOut{Total reward $R(s, a)$}

Compute $R_{\text{eng}}(s, a) \leftarrow$ engagement score (e.g., watch time, scroll depth)\;

\If{emotional state improves from negative to neutral/positive}{
    $R_{\text{emo}}(s, a) \leftarrow \eta$\;
}
\Else{
    $R_{\text{emo}}(s, a) \leftarrow 0$\;
}

Initialize $R_{\text{cause}}(s, a) \leftarrow 0$\;

\ForEach{causal parent $f \in \mathcal{C}$}{
    Compute $\Delta f(s, a) \leftarrow$ improvement in feature $f$\;
    Update $R_{\text{cause}}(s, a) \mathrel{+}= \lambda_f \cdot \Delta f(s, a)$\;
}

Compute total reward: \\
\quad $R(s, a) \leftarrow R_{\text{eng}}(s, a) + \alpha \cdot R_{\text{emo}}(s, a) + \beta \cdot R_{\text{cause}}(s, a)$\;

\textbf{return} $R(s, a)$
\end{algorithm}

\textbf{Training and Execution.}
Each episode simulates a user’s 30-day interaction timeline. On day \( t \), the system observes state \( s_t \), selects an action \( a_t \) via Boltzmann sampling over Q-values, receives the composite reward \( r_t \), and transitions to the next state \( s_{t+1} \) based on the user’s updated emotional state \( e_t \). The Q-values are updated using standard Q-learning principles~\cite{sutton2018reinforcement}.

\textbf{Logging and Convergence.}
After each action, the system logs \((s_t, a_t, r_t, e_t, \pi_t)\) along with the top-K Q-value videos and reward components. Training converges by epoch 8, with rewards plateauing and temperature decay reducing variance in action selection. We use a replay buffer of size 10,000 to reinforce rare but emotionally significant transitions.

\textbf{Hyperparameters.} We tuned the core Q-learning hyperparameters using grid search on a held-out user subset, with the final values summarized in Table~\ref{tab:rl_hyperparams}.

\begin{table}[htbp]
\centering
\small
\setlength{\tabcolsep}{6pt} % shrink column spacing
\caption{RL Agent Hyperparameters for Emotion-Aware Recommendations}
\label{tab:rl_hyperparams}
\resizebox{\linewidth}{!}{%
\begin{tabular}{ll}
\toprule
\textbf{Parameter} & \textbf{Value} \\
\midrule
Learning rate (\(\alpha\)) & 0.1 \\
Discount factor (\(\gamma\)) & 0.95 \\
Emotion transition bonus (\(\eta\)) & 1.0 \\
Causal shaping weight (\(\beta\)) & 0.5 \\
Replay buffer size & 10,000 \\
Training epochs & 10 \\
Exploration strategy & Boltzmann (temperature decay) \\
\bottomrule
\end{tabular}
}
\end{table}

\section{Results and Evaluation}
\label{sec:results}
We assess how ESMR improves user emotion trajectories and engagement outcomes and run Ablation studies for the reward function.
\subsection{\textbf{Causal Discovery for Emotional States}}
\label{sec:causal_dag}

We explore the causal relationships that drive emotional transitions within the ESMR framework. The causal discovery process, detailed in Section~\ref{sec:causal}, identifies key emotional, behavioral, and transitional features associated with target emotions. These features are used to inform the reinforcement learning agent’s reward function. 

Figure~\ref{fig:causal_parents} illustrates the causal parents for Next Day Happy and Next Day Stressed emotional states derived from our causal discovery process. Emotional recovery, for example, is heavily influenced by transitions from Disappointed to Happy, as well as positive behavioral trends such as engagement with content (Refer to Table~\ref{tab:causal_parents_summary} for Top-3 \(\mathcal{P}\)).

We validate the plausibility of the learned DAGs through two mechanisms: (1) qualitative alignment of top causal parents with interpretable behavioral-emotional patterns (e.g., scrolling time and engagement change as strong predictors of \textit{next day stressed} and \textit{next day happy}) consistent with findings from prior literature~\cite{gao2017behavioral, zhang2021emotionaware}; and (2) improved emotional recovery metrics when shaping is applied.

\begin{figure}[htbp]
  \centering
    \includegraphics[width=0.98\columnwidth]{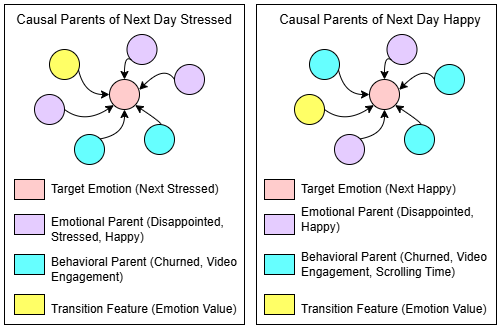}
    \caption{DAG of Causal Parents for Next Day Happy and Next Day Stressed State, highlighting emotional, behavioral, and transitional features identified through causal discovery. Color-coded nodes reveal key drivers of happiness and stress based on user engagement dynamics.}
    \label{fig:causal_parents2}
    \vspace{6pt}   % or 0.2cm

\end{figure}

\subsection{TabTransformer Predictions}
\label{sec:emotion_prediction_results}
\begin{figure}[htbp]
    \centering
    \includegraphics[width=\linewidth]{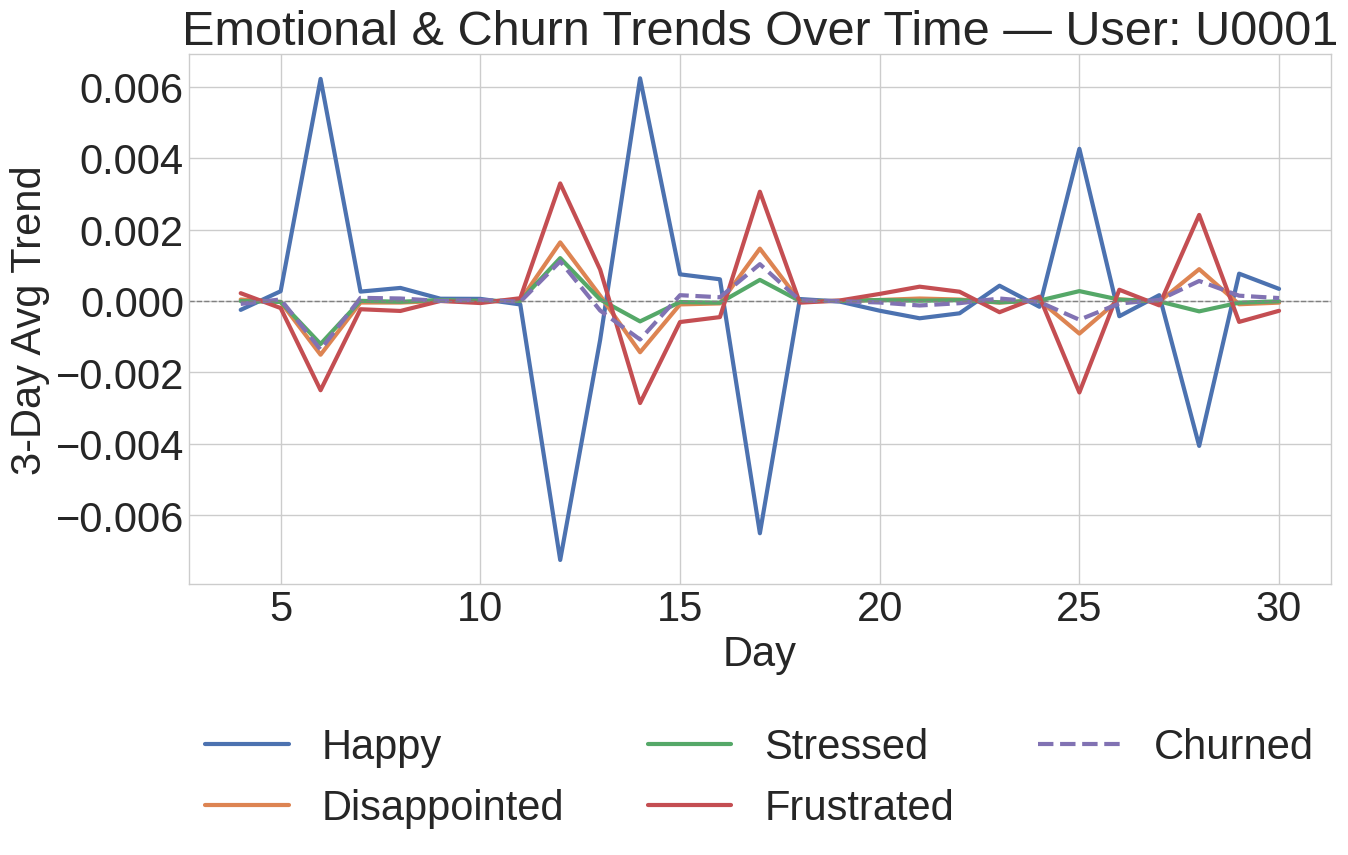}
    \caption{3-day averaged emotional trajectory for user U0001 as predicted by the TabTransformer. Trends reflect affective state evolution in response to engagement and content patterns.}
    \label{fig:emotion-trend}
\end{figure}

Figure~\ref{fig:emotion-trend} shows the 3-day averaged emotional trajectory predicted by the TabTransformer for a representative user (User U0001). The trends reveal how the user’s emotional state evolves in response to content consumption and engagement patterns over time. A 3-day moving average was applied to reduce noise and highlight dominant affective fluctuations. Notably, the user's emotional state exhibits oscillatory patterns across both positive states (e.g., \textit{happy}) and negative states (e.g., \textit{frustrated}), demonstrating the model's ability to capture temporal nuances in emotional transitions.

\begin{figure}[htbp]
    \centering
    \includegraphics[width=\linewidth]{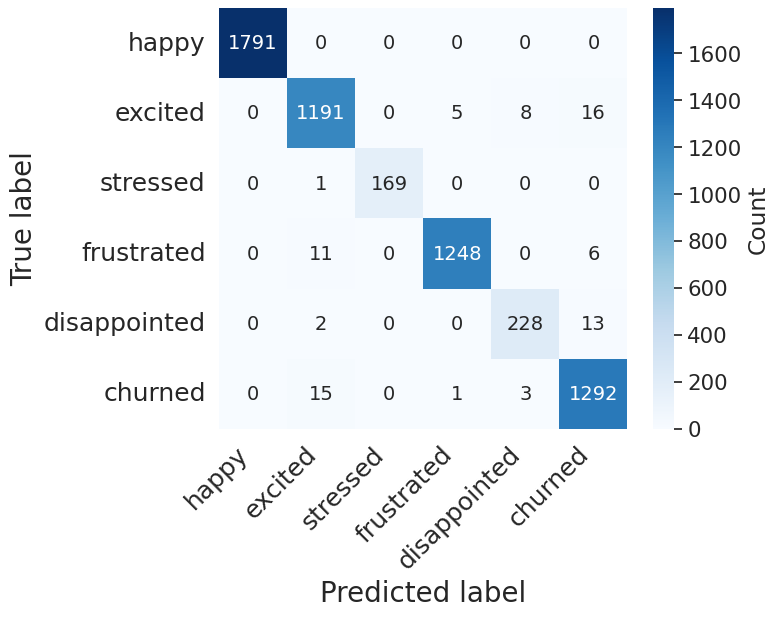}
    \caption{Confusion Matrix}
    \label{fig:-cm}
\end{figure}

Figure~\ref{fig:-cm} presents the confusion matrix for the emotion prediction model, where we achieve a classification accuracy of approximately 98.3\% across six emotional states. The confusion matrix shows strong discriminative ability, particularly for emotionally distinct states like \textit{happy}, \textit{frustrated}, and \textit{churned}. Moderate misclassifications, such as between \textit{excited} and \textit{disappointed}, arise due to shared engagement features like high scrolling and mixed interaction patterns. These insights confirm the model's strong classification power and its ability to distinguish key emotional states effectively.

\iffalse
\begin{table}[htbp]
\centering
\small
\caption{Evaluation metrics and model variants used in our experiments.}
\label{tab:eval-setup}
\begin{tabular}{@{}p{3.6cm}p{4.4cm}@{}}
\toprule
\textbf{Metric / Model} & \textbf{Description} \\
\midrule
\multicolumn{2}{l}{\textit{Evaluation Metrics}} \\
\midrule
\textbf{Recovery Time} & Days until reaching stable positive or neutral emotional states. \\
\textbf{Emotion Valence} & Mean affective value over the 30-day simulation. \\
\textbf{Negative Emotion Days} & Number of days in negative dominant emotional states. \\
\textbf{Emotion Volatility} & Standard deviation of daily dominant emotions. \\
\textbf{Bounce Rate} & Percentage of users oscillating between extreme affective states. \\
\textbf{Engagement Score} & Average daily behavioral reward (e.g., watch time). \\
\textbf{Causal/Threshold Bonus} & Cumulative bonus from shaping during RL activation. \\
\midrule
\multicolumn{2}{l}{\textit{Model Variants}} \\
\midrule
\textbf{LightGBM-only} & Engagement-based policy without any shaping. \\
\textbf{Emotion-only RL} & RL with only emotion shaping. \\
\textbf{Engagement-only RL} & RL with only engagement shaping. \\
\textbf{ESMR (Hybrid)} & Full RL with emotion, engagement, and causal shaping. \\
\bottomrule
\end{tabular}
\end{table}
\fi

\subsection{Emotional Recovery and Stability}
\label{sec:emotional_recovery}

We evaluate the performance of ESMR in promoting smoother affective transitions and helping users recover from early negative emotional states. Instead of only focusing on final emotion labels, we assess emotional recovery as a temporal process, emphasizing volatility, streak duration, and positive shifts over time.

ESMR reduces the average number of days spent in negative emotional states from 5.22 (LightGBM) to 2.14 and increases the proportion of users reaching \texttt{happy} at least once to 97.7\%. Furthermore, 34.9\% of users conclude with \texttt{happy} as their final dominant emotion, and 35.4\% achieve full recovery from initial negative states.

To measure emotional stability, we compute two indicators: (i) \textbf{volatility}, defined as the standard deviation of daily dominant emotions, and (ii) \textbf{bounce rate}, the percentage of users oscillating between high-intensity positive and negative states (e.g., \texttt{excited}~$\leftrightarrow$~\texttt{angry}) more than twice. ESMR reduces volatility by 37.8\% and bounce rate from 57.5\% to 0.4\%.

\begin{figure}[htbp]
  \centering
  \includegraphics[width=0.47\textwidth]{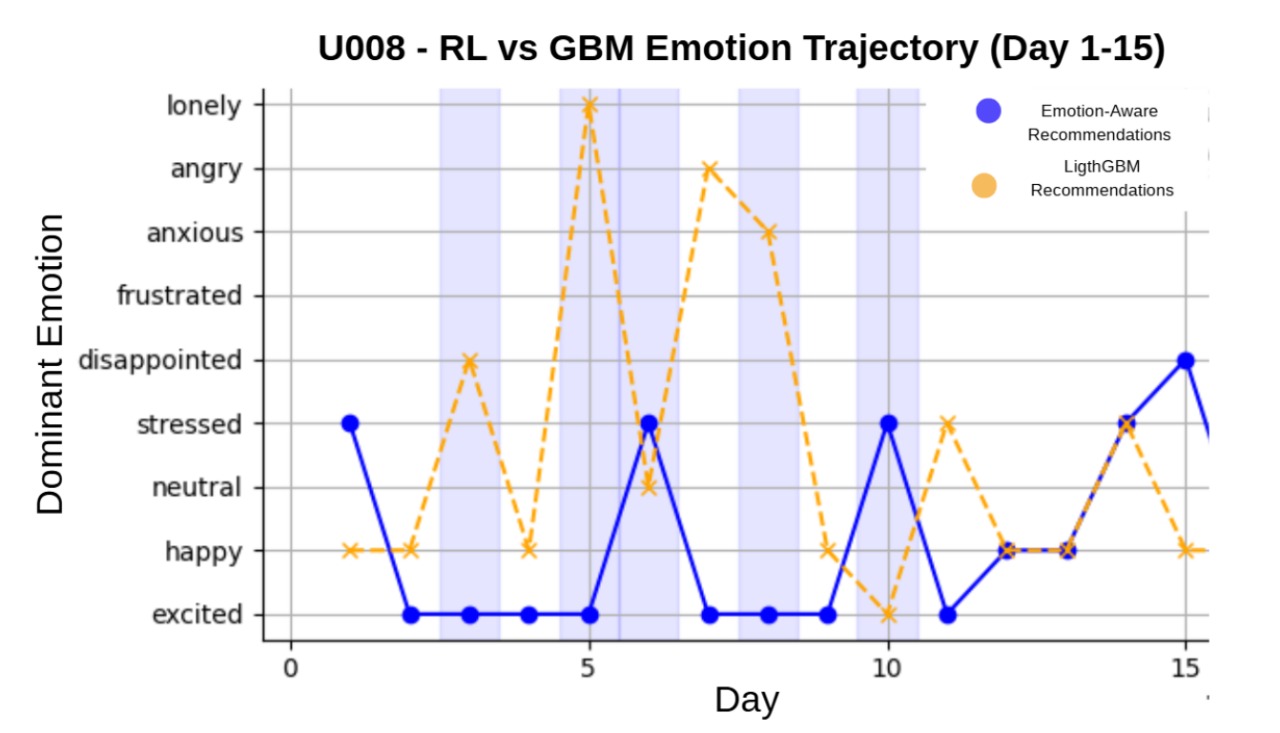}
  \includegraphics[width=0.47\textwidth]{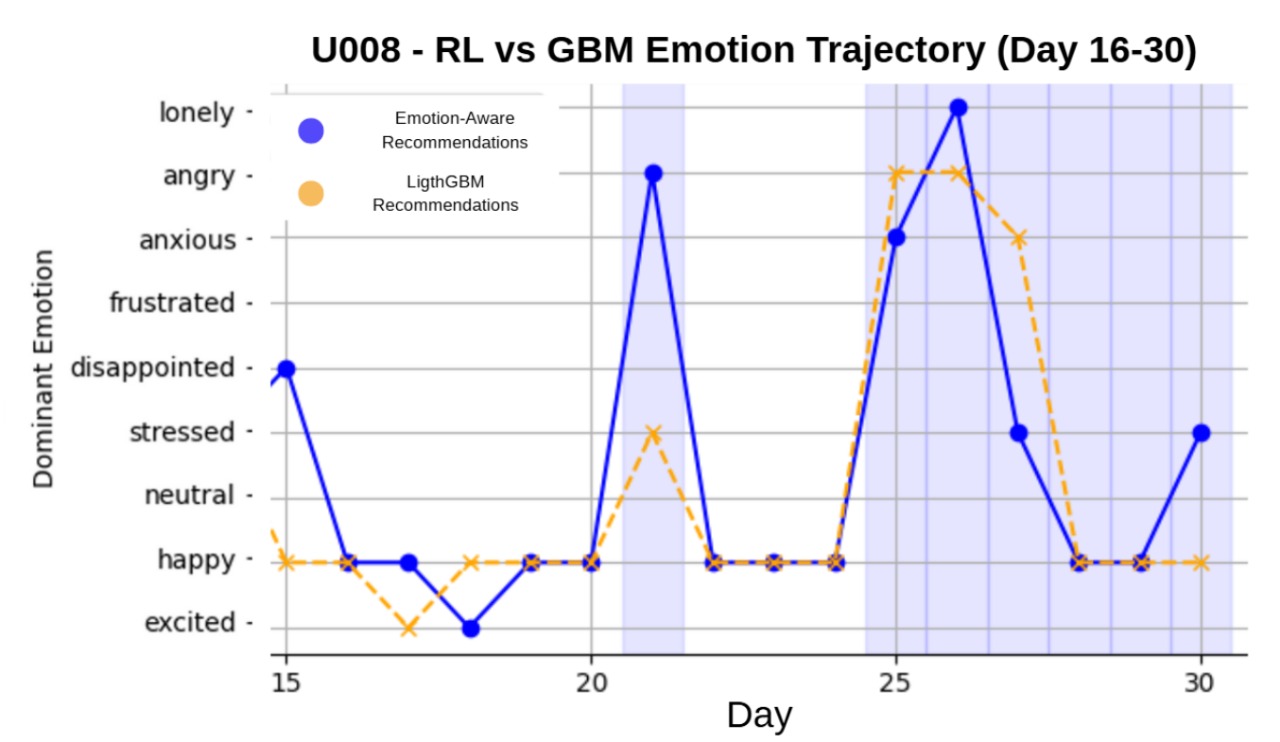}
  \caption{Dominant emotion trajectory for User 008 over 30 days. RL-activated days (shaded) yield smoother, more emotionally stable transitions than the LightGBM-only baseline.}
  \label{fig:user-u008-trajectories}
\end{figure}

Figure~\ref{fig:user-u008-trajectories} illustrates the \textbf{30-day emotional trajectory} of a representative user (U008). Under ESMR, the user transitions more smoothly toward stable positive states, with fewer erratic swings, compared to the LightGBM-only baseline.

ESMR achieves significant emotional improvements while maintaining high engagement, with only a 2.25\% reduction in average engagement reward compared to LightGBM (1402.8 vs. 1435.1). This demonstrates that selectively activating hybrid RL models can promote emotionally intelligent content sequencing and facilitate faster recovery from distress, without sacrificing engagement.

\subsection{Ablation Studies}
\label{sec:ablation}

To isolate the contributions of emotion and engagement shaping within ESMR, we perform ablation studies by selectively disabling one reward component at a time. All variants use the same simulation setup, Q-learning agent, and recommendation space. Metrics are computed over all users and logged episodes.

Table~\ref{tab:ablation-results} compares the full Hybrid policy (ESMR) with two ablations: (i) emotion shaping disabled (Engagement-only), and (ii) engagement shaping disabled (Emotion-only).

\begin{table}[htbp]
\centering
\small
\caption{Ablation study comparing ESMR with Emotion-only and Engagement-only variants.}
\label{tab:ablation-results}
\resizebox{\columnwidth}{!}{%
\begin{tabular}{@{}lccc@{}}
\toprule
\textbf{Metric} & \textbf{ESMR} & \textbf{Emotion = 0} & \textbf{Engagement = 0} \\
\midrule
Emotion Valence & \textbf{-0.006} & -0.029 & -0.029 \\
Negative Emotion Days & \textbf{14.98} & 15.45 & 14.98 \\
Recovery Time & \textbf{7.82 days} & 7.74 days & 7.82 days \\
Engagement Score & \textbf{0.188} & 0.181 & 0.188 \\
\bottomrule
\end{tabular}%
}
\end{table}

Disabling emotion shaping results in worse emotional outcomes across all metrics: lower emotion valence, more negative emotion days, and slower recovery times. On the other hand, disabling engagement shaping preserves affective stability, confirming that ESMR still maintains engagement comparable to the full model.

These results underscore the complementary roles of emotion and engagement shaping. While engagement-only RL maintains behavioral metrics, the full ESMR framework significantly improves emotional trajectories through balanced shaping, highlighting the importance of emotion-aware content recommendations.

\section{Conclusion}

This work introduces ESMR, a hybrid reinforcement learning framework for emotion-aware social media recommendations that integrates engagement dynamics, affective feedback, and causal shaping. Using a 30-day user simulation based on real behavioral patterns, we show that ESMR significantly reduces emotional volatility and negative streaks while maintaining high engagement. Ablation studies confirm the complementary roles of emotion and engagement shaping, and per-user analyses highlight stable, supportive emotional trajectories. Importantly, our causal discovery process reveals that user emotions are emergent, driven by content categories, behavioral features, and emotion-labeled video input, validating our approach of using video-level emotion to influence user affect.

While this study relies on simulated users due to the lack of large-scale real-world emotion–engagement logs, the system is fully reproducible, with open data generation, model training, and evaluation pipelines. These findings suggest that emotion-aware recommendation systems, guided by causally grounded and selectively triggered RL, can offer a promising path toward healthier, more adaptive platforms. Future work will explore real-time adaptation, human-in-the-loop evaluations, and LLM-driven narrative shaping to further personalize recovery-oriented recommendations, highlighting the importance of prioritizing long-term user well-being alongside engagement.

\section{Ethical Statement}
While our system aims to promote emotional well-being through content curation, we acknowledge the ethical risks of unintended emotional manipulation. Future deployments must incorporate user consent, transparency, and safeguards such as user controls and explainable recommendations.

\bibliographystyle{unsrt}   % or plainnat, abbrvnat
\bibliography{references}
\end{document}